\documentclass[preprint,proceedings]{rmaa_adapted}

\usepackage{amsmath}
\usepackage{graphicx}
\usepackage{psfig}
\usepackage{natbib}
\usepackage{makeidx}
\usepackage{sidecap}

\usepackage{paralist}

\usepackage{psfrag,color}

\SetYear{2003}
\SetConfTitle{Revista Mexicana de Astronom\'ia y Astrof\'isica}

\title{The Primordial Binary Population in OB Associations} 

\author{
  M. B. N. Kouwenhoven\altaffilmark{1}, 
  A. G. A. Brown\altaffilmark{2}, 
  A. Gualandris\altaffilmark{1}, 
  L. Kaper\altaffilmark{1},
  S. F. Portegies Zwart\altaffilmark{1,3}, 
  and H.~Zinnecker\altaffilmark{4}. 
}
\altaffiltext{1}{Astronomical Institute ``Anton Pannekoek'', Amsterdam, The Netherlands.}
\altaffiltext{2}{Leiden Observatory, The Netherlands.}
\altaffiltext{3}{Institute for Computer Science, Amsterdam, The Netherlands.}
\altaffiltext{4}{Astrophysikalisches Institut Potsdam, Germany.}

\shortauthor{Kouwenhoven et al.}
\shorttitle{The Primordial Binary Population}

\fulladdresses{
\item Kouwenhoven, Gualandris, Kaper and Portegies Zwart: Astronomical Institute ``Anton Pannekoek'', University of Amsterdam, Kruislaan 403, 1098 SJ, Amsterdam, The Netherlands (\email{kouwenho, alessiag, lexk, spz@astro.uva.nl}).
\item Brown: Leiden Observatory, University of Leiden, P.O. Box 9513, NL-2300 RA Leiden, The Netherlands (\email{brown@strw.leidenuniv.nl}).
\item Portegies Zwart: Institute for Computer Science, University of Amsterdam, Kruislaan 403, 1098 SJ, Amsterdam, The Netherlands (\email{spz@astro.uva.nl}).
\item Zinnecker: Astrophysikalisches Institut Potsdam, An der Sternwarte 16, D-14482, Potsdam, Germany (\email{hzinnecker@aip.de}).
}

\listofauthors{Kouwenhoven, M. B. N., Brown, A. G. A., Gualandris, A., Kaper, L., Portegies Zwart, S. F. \& Zinnecker, H.}
\indexauthor{Kouwenhoven, M. B. N.}
\indexauthor{Brown, A. G. A.}
\indexauthor{Gualandris, A.}
\indexauthor{Kaper, L.}
\indexauthor{Portegies Zwart, S. }
\indexauthor{Zinnecker, H.}

\abstract{We present the first results of our adaptive optics survey of 200 (mainly) A-type stars in the nearby OB association Sco~OB2, which we will use, together with literature data and detailed simulations of young star clusters, to determine the primordial binary population.}

\addkeyword{Binaries}
\addkeyword{OB associations}
\addkeyword{Individual: Scorpius~OB2}

\begin{document}
\maketitle

\section{Introduction} \label{sec: Introduction}

For understanding the process of star formation it is essential to know how many stars are formed in binary and multiple systems, as a function of environment and binary parameters. This requires a characterization of the primordial binary population (PBP), which we define as the population of binaries that is present just after star formation has ceased, but before dynamical and stellar evolution have significantly altered its characteristics. From a modeling point of view, this occurs at the time when star formation simulations such as those of Bate et~al.\ (2003) end, and when N-body simulations such as those of Portegies Zwart et al.\ (2001) begin.

Characterizing the PBP is also important for understanding the formation and evolution of OB associations, the origin of the field star population and OB runaway stars, as well as the production and evolution of binary stars. For reviews on different aspects of this issue, we refer to Blaauw (1991), Mermilliod (2001) and Zinnecker (2003). 

Sco~OB2 is a young (5 -- 20 Myr), low-density ($\approx$ 0.1 M$_\odot$ pc$^{-3}$) stellar clustering and therefore the binary population is expected to be relatively unevolved. Moreover, Sco~OB2 is nearby (120 -- 140 pc), has a fully populated IMF and has been cleared of the gas out of which the stars have formed, which facilitates a detailed and comprehensive study. This makes Sco~OB2 an ideal place to determine the PBP.

We carried out an adaptive optics survey of 200 (mainly) A-type stars in the three subgroups Upper Scorpius (US), Upper Centaurus Lupus (UCL), and Lower Centaurus Crux of Sco~OB2 (see Table~\ref{tab: statistics}). These stars were selected from the \textit{Hipparcos} membership study of de~Zeeuw et al.\ (1999).

\section{Observations and Data Reduction} \label{sec: Observations}

The observations were done in the near-infrared with the adaptive optics instrument ADONIS on the ESO 3.6m telescope on La Silla. The adaptive optics observations bridge the observational gap between the wide visual and close spectroscopic binaries, and infrared observations have the advantage over optical observations that the luminosity contrast between primaries and companions is strongly reduced.

The camera field of view is $12.8 \times 12.8$~arcsec. Each star is observed four times in order to maximize the available field of view for study ($\approx 19.2 \times 19.2$~arcsec). All stars are observed in the $K_S$ band, while several stars are also observed in the $J$ and $H$ bands.

After selecting the best images we applied a standard data reduction procedure and find that the typical Strehl ratios are $\approx 5\%$, $10\%$ and $25\%$, in $J$, $H$ and $K_S$, respectively.  The instrumental magnitudes of all visible components were determined using \textit{Starfinder} (Diolaiti et~al.\ 2000), and were calibrated to the 2MASS system. We find a total of 151 stellar components in the fields around the 200 target stars. For each visible component the projected distance, position angle, and relative magnitude are calculated.

\section{Binary statistics}

\begin{figure}[!t]
  \includegraphics[width=\columnwidth]{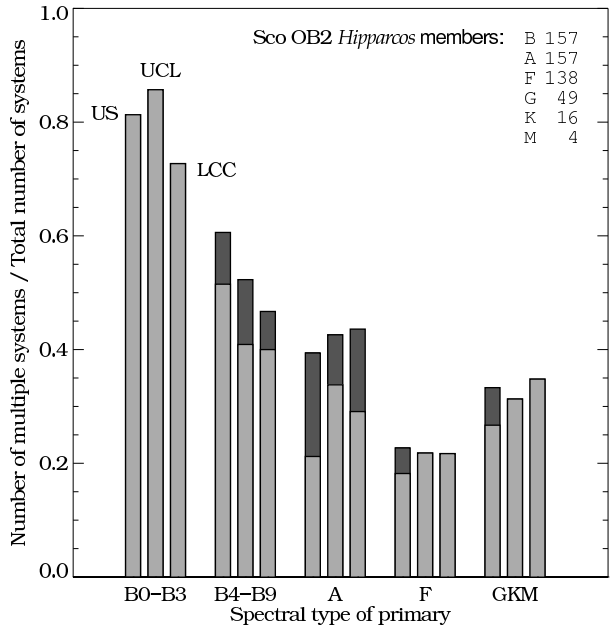}
  \caption{The fraction of stellar systems which are multiple, vs. the spectral type of the primary, for the three subgroups of Sco~OB2 (only \textit{Hipparcos} members). The 47 new companion stars that are presented in this paper are indicated with the dark gray extensions of the bars.}
  \label{fig: mult}
\end{figure}

Since for most stars we only have $K_S$ band observations, we are unable to distinguish between companions and background stars using the color-magnitude diagram. However, secondaries must lie on the main sequence or above. For this preliminary study we follow Shatsky \& Tokovinin (2002) and consider all components fainter than $K = 12$ as background stars. Applying this criterion to our data we find 73 probable background stars, and 78 candidate companion stars. A thorough study of the detection limits is required to determine if we missed faint and red brown dwarfs with this selection criterion.

The list of candidate companions thus obtained is cross-checked with spectroscopic, astrometric and visual binary data in literature, including the complementary study of B-type stars in Sco~OB2 by Shatsky \& Tokovinin (2002). This results in 47 new candidate companions (14 in US, 20 in UCL, and 13 in LCC). The results are displayed in Figure~\ref{fig: mult} and Table~\ref{tab: statistics}, where the number of single stars, binaries, triples, and multiples of order larger than 3 are denoted by $N_s$, $N_b$, $N_t$ and $N_m$, respectively. The distances (in pc) and ages (in Myr) for the subgroups are taken from de Zeeuw et~al.\ (1999).

Brown (2001) pointed out that the observed number of multiple systems with A- and late B-type primaries is relatively low due to observational biases. Part of this selection effect is now removed with our new adaptive optics observations.

\section{Future Work}

The ultimate goal of this project is to determine the PBP. We will achieve this by combining the results described here with available literature data and detailed numerical models of OB associations. These models are N-body simulations including state of the art stellar and binary evolution (see Portegies Zwart et~al.\ 2001). Simulated observations of model OB associations will be produced to characterize in detail the selection effects for the different binary surveys.

These cluster simulations will also be used to investigate the impact of dynamical and evolutionary effects that may have altered the binary population over the lifetime of Sco~OB2. Combined with the knowledge about the selection effects, the primordial binary population can then be reconstructed, giving binary characteristics as a function of environmental properties.

\begin{table}[!t]
  \centering
  \setlength{\tabcolsep}{0.7\tabcolsep}
  \caption{Multiplicity in Sco~OB2\tabnotemark{(a)}} 
  \begin{tabular}{l cc cccc cc}
    \toprule
          & $D$  & Age  & $N_s$ & $N_b$ & $N_t$ & $N_m$ & $f_s$\tabnotemark{(b)} & $f_c$\tabnotemark{(c)} \\
    \midrule
    US    & 145  & 5    & 63    & 46    & 7     & 3     & 0.47  & 0.67  \\
    UCL   & 140  & 13   & 128   & 70    & 18    & 4     & 0.42  & 0.62  \\
    LCC   & 118  & 10   & 112   & 55    & 12    & 0     & 0.37  & 0.56  \\
    \midrule
    all   &      &      & 303   & 171   & 37    & 7     & 0.42  & 0.62  \\
    \bottomrule
  \end{tabular}
  \setlength{\tabnotewidth}{1.9\textwidth}
  \begin{tabular}{l}
    \tabnotetext{(a)}{Only \textit{Hipparcos} members are considered.}
    \tabnotetext{(b)}{$f_{\rm s} = (N_b+N_t\dots) \ /\ (N_s+N_b+N_t+\dots)$, } 
    \quad the fraction of multiple systems. \\
    \tabnotetext{(c)}{$f_{\rm c} = (2N_b+3N_t\dots) \ /\ (N_s+2N_b+3N_t+\dots)$, } 
    \quad the fraction of stars in multiple systems. \\
  \end{tabular}
  \label{tab: statistics}  
\end{table}

\end{document}